\documentstyle[12pt,epsf]{article}
\title{Radiative Dileptonic Decays of B Mesons}
\author{ Gad Eilam, Cai-Dian  L\"{u} and Da-Xin Zhang\\
{\small Physics Department, Technion- Israel Institute of Technology, 
  32000 Haifa, Israel}}

\date{}
\begin{document}
\maketitle
\begin{picture}(0,0)(0,0)
\put(330,270){{\large hep-ph/9606444}}
\put(330,250){{\large TECHNION-PH-96-12}}
\put(330,230){{\large June, 1996}}
\end{picture}

\begin{abstract}
We investigate  the radiative  dileptonic decays $B_s(B_d)\to\gamma 
l^+l^-$ within the standard model. Using the constituent  quark model, 
the branching ratios turn out to be around $5\times 10^{-9}$ for  
$B_s\to \gamma \mu^+\mu^- $ and around $6\times 10^{-10}$ for $B_d \to 
\gamma \mu^+\mu^- $, with slightly larger values for $B_s(B_d) \to 
\gamma e^+e^- $. The differential rate as a function of  the dilepton 
invariant mass is given. The possibility of using these processes to 
determine the decay constants of $B_s$ or $B_d$ is discussed. 
\end{abstract}
\newpage

\section{Introduction}
Rare decays induced
by flavor changing neutral currents can be used 
as tests of the Standard Model (SM), and are sensitive to new physics.
It is expected that at future  B factories and fixed target machines,
 other rare decay channels 
of the bottom quark \cite{smrev} will be discovered in addition to
the observed $b\to s\gamma$ transition\cite{cleo}. 
These processes also offer useful information
for  extracting the fundamental parameters of the SM, 
 such as  $|V_{ub}|$ \cite{wise}.

Rare decays can also serve as  alternative channels to measure
some elementary hadronic parameters. 
For instance, the decay constants $f_{B_q}$, $q=s,d$ can be extracted from 
$B_{q} \to \gamma \nu\bar\nu $ \cite{nunubar}.
In the present work,
we investigate the possibility of using the  channel, 
 $B_{q} \to \gamma l^+l^-$ to determine the decay constants.
Pure leptonic decays of heavy pseudoscalar mesons
into light lepton pairs are helicity suppressed, their branching ratios are 
of the order of $10^{-9}$ for $B_s\to\mu^+\mu^-$, and $10^{-14}$ for $
B_s\to e^+e^-$ \cite{stone}.
This makes it difficult to  determine $f_{B_s}$ from these processes.
For $B_d$ the situation is  even worse due to the smaller CKM angle.
If a photon is emitted in addition to the lepton pair, 
then the mechanism of helicity suppression will not hold any longer
and larger branching ratios are expected. Although
the process $B_s\to\tau^+\tau^-$ is free from helicity
 suppression and its branching ratio
is around $8\times 10^{-7}$ in the SM \cite{buras},
it is likely to be compatible with the decays into lepton pairs
 only when its efficiency is better than $10^{-2}$.

In Section 2 the relevant effective Hamiltonian will be given in the SM, and
the constituent quark model will be used to give the numerical predictions.
Finally,  we make some comments in Section 3.

\section{Model calculations }

The most important contributions to $B_q \to \gamma l^+l^-$
($l=e, \mu$) stem from the  effective Hamiltonian
which induces  the pure leptonic processes $B_q \to  l^+l^-$.
Let us start with the short distance contributions to the latter one,
which include  box, $Z$ and photon penguin 
diagrams. The Feynman diagrams are displayed in Fig.1. 
The QCD corrected effective Hamiltonian in the 
SM is \cite{bsll,mis}:
\begin{equation}
{\cal H}_{eff} =\frac{\alpha G_F }{\sqrt{2}\pi}
 V_{tb}V_{tq}^* \left[ \left (C_9^{eff}\bar q
\gamma^\mu P_L b +\frac{2C_7m_b}{p^2} \bar q \not \! p \gamma^\mu 
P_R b\right ) \bar l\gamma_\mu l \; + C_{10} (\bar q\gamma^\mu P_L b)
\bar l\gamma_\mu \gamma_5 l\right],\label{1}
\end{equation}	 
with $P_L=(1-\gamma_5)/2$, $P_R=(1+\gamma_5)/2$, and $p=p_{+} +p_{-}$ is the 
momentum of the lepton pair. 
The QCD corrected Wilson coefficients $C_7(m_b)$, $C_9^{eff}(m_b)$, and 
$C_{10}(m_b)$ are given by Misiak \cite{mis}.

Because of the lightness of the leptons $e$ and $\mu$,
the pure leptonic processes $B_q\to l^+l^-$ are suppressed by helicity. 
If a photon line is attached to any of the charged lines in Fig.1, the 
situation will be different:  helicity suppression is overcome. 
When the photon  is attached to  internal  lines, 
there will be a suppression factor 
of $m_b^2/M_W^2$  in the Wilson coefficient,
since the resulting operators(dimension-8 ) are two  orders higher 
in dimensions than the usual ones(dimension-6).

One may expect that the additional two diagrams displayed in Fig.2 (a) and (b)
might also contribute to the decay $  B_q\to \gamma l^+l^-$.
However, our calculation shows that the contribution of 
Fig.2(b) is quite small and its influence can be neglected in the numerical 
results.
Fig.2(a) involves a singularity, since the intermediate $s$ quark can be
on its mass shell. 
The influence of this singularity is an artifact of the constituent quark 
model employed here. If we calculate the contribution of Fig. 2(a) to the
 differential decay rate anywhere except near the singularity we find it to be
negligible ( at most a few percent). Therefore we neglect it altogether.

There are also long distance contributions to $B_q \to l^+l^- \gamma$. 
For instance, in the $B_s$ decay, there are 
 cascade processes at the hadronic level:
\begin{equation}
B_s\to \phi (\phi', J/\psi...) \gamma,~~ \phi (\phi', J/\psi...) \to l^+l^-.
\end{equation}
These involve the on-shell $\phi,\phi', J/\psi...$  and are similar to
$b\to s J/\psi \to s l^+l^-$ \cite{vmd}.
They are estimated to be of order of $10^{-9}$, which is about the same 
order of the short distance contribution.
 All these long distance effects, which are shown as sharp peaks in the 
invariant mass spectrum of the lepton pair, will not be included in the 
following.

Now using a simple  constituent quark model
(see, for example, \cite{cheng}), we calculate the amplitude for diagrams
 with photon emitted from external fermion lines.
First, for diagrams with photon attached to external lepton lines,
we get the amplitude proportional to:
\begin{equation}
\bar q \gamma_\mu b_L \epsilon _\nu \bar l \left[ \gamma^\nu
\frac{\not\! p_- + \not\! p_\gamma +m_l}{(p_- +p_\gamma)^2-m_l^2} 
\gamma^\mu (C_9^{eff}+C_{10}\gamma_5 )- \gamma^\mu (C_9^{eff}+C_{10}\gamma_5 )
\frac{\not\! p_+ + \not\!  p_\gamma -m_l}{(p_+ +p_\gamma)^2-m_l^2} 
\gamma^\nu \right] l.\label{4}
\end{equation}
Using the decay constant definition:
\begin{equation}
<0|\bar q \gamma^\mu \gamma_5 b|B> =- f_{B_q} p_B^\mu,\label{dd}
\end{equation}
we can easily calculate the results of eqn.(\ref{4}). Note that,
there is no contribution from the operator $O_7$ in this circumstance,
since the definition of the decay constant shows that:
\begin{equation}
<0|\bar q \sigma^{\mu\nu} P_R b|B>=0.
\end{equation}

For the other two diagrams with photon emitted from the external quark 
lines ($b$ or $q$), the amplitude is proportional to:
$$
\epsilon _\nu \bar q \left[ \gamma^\nu 
\frac{ \not\! p_\gamma - \not\!  p_q +m_q}{(p_q \cdot p_\gamma)} 
\gamma^\mu P_L +P_R  \gamma^\mu
\frac{\not\! p_b - \not\! p_\gamma +m_b}{(p_b \cdot p_\gamma)} 
\gamma^\nu  \right] b \left[ C_9^{eff} \bar l \gamma_\mu l 
+C_{10} \bar l \gamma_\mu \gamma_5 l) \right]
$$
\begin{equation}
+2 \frac{C_7 m_b }{p^2} \epsilon _\nu \bar q \left[
P_R \not\! p \gamma^\mu 
\frac{\not\! p_b - \not\! p_\gamma +m_b}{(p_b \cdot p_\gamma)}
\gamma^\nu + \gamma^\nu 
\frac{\not\! p_\gamma - \not\! p_q +m_q}{(p_q \cdot p_\gamma)} 
\not\! p \gamma^\mu P_R \right] b \;(\bar l \gamma_\mu l ).
\label{4.1}
\end{equation}
In the constituent quark model, $p_q^\mu=(m_q /m_B)p_{B_q}^\mu$, and
$p_b^\mu=(m_b /m_B)p_{B_q}^\mu$. Then applying eqn.(\ref{dd}),
 neglecting terms suppressed by $m_q/m_b$, $q=d,s$, we get:
\begin{equation}
{\cal A} = C \left[  i
 \epsilon_{\alpha \beta \mu\nu}
\epsilon_\gamma^\alpha p_\gamma^\beta p_{B_q}^\nu
+ 
(p_{\gamma \mu}\epsilon_{\gamma\nu}- p_{\gamma\nu} 
\epsilon_{\gamma\mu})p_{B_q}^\nu \right]
\left[\left(C_9^{eff}-\frac{2C_7 m_{B_q}^2}{p^2} \right)\bar l
 \gamma ^\mu l+C_{10} \bar l\gamma^\mu \gamma_5 l \right],\label{5}
\end{equation}
with 
$$C\equiv \frac{e \alpha G_F f_{B_q}m_{B_q}}{12\sqrt{2}\pi m_q
( p_{B_q} \cdot p_\gamma)}V_{tb}V_{tq}^*.$$
 After squaring the total decay amplitude, we find that the contribution
of diagrams with photon attached to the lepton 
lines (eqn.(\ref{4}) ) is proportional to $m_l$, which is negligible in
numerical calculations. The  main contribution to $  B_q  \to \gamma 
l^+l^-$ is thus from the square of eqn.(\ref{5}).
 Performing the phase space integration over one of the two Dalitz 
variables,  we get the 
differential decay width versus $ s= (p_++p_-)^2/m_{B_q}^2$:
\begin{equation}
\frac{d\Gamma}{d  s} =\frac{ \alpha^3 G_F^2 f_{B_q}^2m_{B_q}^5 }
{3456\pi^4 m_q^2} |V_{tb}V_{tq}^*|^2
(1- s)  s \left(C_{10}^2 +\left|C_9^{eff}-\frac{2}{ s}
C_7\right|^2\right).\label{6}
\end{equation}
We see that the differential rate is proportional to $(f_{B_q}/m_q)^2$.
Using $\alpha=1/137$, $m_s=0.51$ GeV, 
 $m_t=176$ GeV and $|V_{tb} V_{ts}^*|=0.04$, 
the differential decay rate as a function of $ s$ is displayed in Fig.3,
which shows that the contribution from soft photons, corresponding to large
$ s$ region, is negligibly small.

After integration, we also get:
\begin{eqnarray}
\Gamma (B_s \to \gamma e^+e^- )&=&3.0\times 10^{-21} \times
\left(\frac{f_{B_s} }{0.2{\rm GeV} }\right)^2~{\rm GeV}. \nonumber\\
\Gamma (B_s \to \gamma \mu^+\mu^- )&=&2.2\times 10^{-21} \times
\left(\frac{f_{B_s} }{0.2{\rm GeV}}\right)^2~{\rm GeV}. \label{9}
\end{eqnarray}
For $B_d$ meson decay, we take  $|V_{tb}V_{td}^*|=0.01$ 
and $m_d=0.35$ GeV.  
The decay widths are then:
\begin{eqnarray}
\Gamma (B_d \to \gamma e^+e^- )&=&3.6\times 10^{-22} \times
\left(\frac{f_{B_d} }{0.2{\rm GeV}}\right)^2~{\rm GeV}, \nonumber\\
\Gamma (B_d \to \gamma \mu^+\mu^- )&=&2.7\times 10^{-22} \times
\left(\frac{f_{B_d} }{0.2{\rm GeV}}\right)^2~{\rm GeV}. \label{10}
\end{eqnarray}
If the lifetimes are taken as $\tau(B_s)=1.34\times 10^{-12} s$,
 $\tau(B_d)=1.50\times 10^{-12} s$ \cite{pdg}, and if $f_{B_q}=200$MeV is used,
they correspond to the branching ratios:
\begin{eqnarray}
B (B_s \to  e^+e^- \gamma)&=&6.2\times 10^{-9} , \nonumber\\
B (B_s \to  \mu^+\mu^- \gamma)&=&4.6\times 10^{-9}, \\
B (B_d \to  e^+e^- \gamma)&=&8.2\times 10^{-10}, \nonumber\\
B (B_d \to  \mu^+\mu^- \gamma)&=&6.2\times 10^{-10}.\nonumber 
\end{eqnarray}

Note that the branching ratios we get for radiative decays are just 
the same order of the pure leptonic decay $B_q\to \mu^+\mu^-$ . 
The decay rates for  pure leptonic decays are: 
\begin{equation}
\Gamma(B_q\to l^+l^-) =\frac{ \alpha^2 G_F^2 f_{B_q}^2m_{B_q}m_l^2}
{16\pi^3} |V_{tb}V_{tq}^*|^2  C_{10}^2.
\end{equation}
Using the same parameters as in the radiative decay, the numerical 
results are:
\begin{eqnarray}
B (B_s \to  e^+e^- )&=&6.1\times 10^{-14} , \nonumber\\
B (B_s \to  \mu^+\mu^- )&=&2.6\times 10^{-9}, \\
B (B_d \to  e^+e^- )&=&4.2\times 10^{-15}, \nonumber\\
B (B_d \to  \mu^+\mu^- )&=&1.8\times 10^{-10}.\nonumber 
\end{eqnarray}
So for the $e^+e^-$ channel, the radiative decay dominates over 
the dileptonic one, while
for the $\mu^+\mu^-$ channel, the branching ratio of the radiative decay is 
a little larger than that for the pure leptonic decay.

\section{Conclusions}

We predict the branching ratios in the SM for $B_s \to \gamma 
l^+l^- $ to be around $5\times 10^{-9}$ for  
$B_s\to \gamma \mu^+\mu^- $ and around $6\times 10^{-10}$ for $B_d \to 
\gamma \mu^+\mu^- $, with slightly larger values for $B_s(B_d) \to 
\gamma e^+e^- $.
With these predictions, they will hopefully be detected at LHC-B. 
thus provide alternative
channels for measuring $f_{B_s}/m_s$ ($f_{B_d}/m_d$). In LHC-B, 
approximately $6\times 10^{11}$  ($2\times 10^{11}$) $B_d$ ($B_s$) mesons 
are expected per year, therefore 
there is a good chance of observing the decays considered here. 

We conclude with comments on the experimental extractions of the decay 
constants. In the measurements of the pure leptonic decays,
processes with additional soft photons emitted (say, photons with energy 
less than 50 MeV) might  contribute due to the inability of the detector 
to separate them. From our calculation,
we find that these soft photon processes are not important 
in the radiative dileptonic decays, see Fig.3. Since the efficiency of 
detecting photons may be low, it will be useful for 
experiments to have the sum of the radiative decay and the pure leptonic 
processes:
\begin{eqnarray}
B(B_s\to\mu^+\mu^-+\gamma\mu^+\mu^-)&=&7.3\times 10^{-9},\nonumber\\
B(B_d\to\mu^+\mu^-+\gamma\mu^+\mu^-)&=&8.1\times 10^{-10}.
\end{eqnarray}
The total $\mu^+\mu^-(\gamma)$ branching ratios for $ B_s$ and $B_d$ decay
are almost the same as that of the $e^+e^- \gamma$ case. 

\section*{Acknowledgement}

The research of D.-X. Z. is supported in part by Grant 5421-3-96
from the Ministry of Science and the Arts of Israel.
The research of G.E. is supported in part by the Israel Science Foundation.
G.E. would like to thank D. Atwood, T. Nakada, A. Soni and the late 
R. Mendel, for very helpful discussions.

\section*{Figure Captions}
\noindent

Fig.1 Feynman diagrams in standard model for $b\bar q\to l^+l^-$.

Fig.2  Additional diagrams which contribute to 
$B_s\to\gamma l^+l^-$.
The relevant penguin operators are $O_7$ (black dot).

Fig.3 Differential decay rates of $B_s \to \gamma l^+l^-$ versus
$ s=(p_++p_-)^2/m_{B_s}^2$.


\newpage
\begin{figure}
\centerline{\epsffile{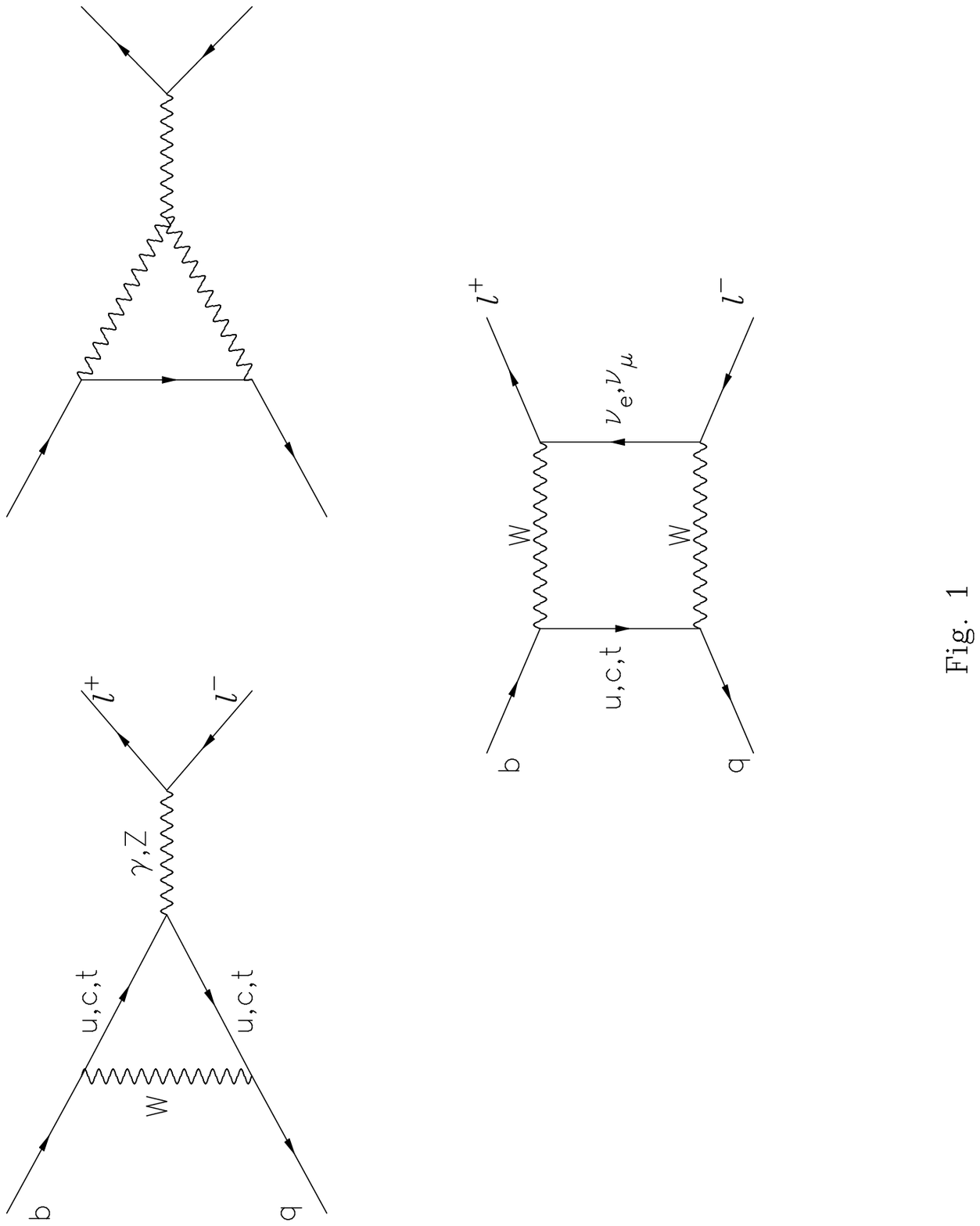}}
\end{figure}

\begin{figure}
\centerline{\epsffile{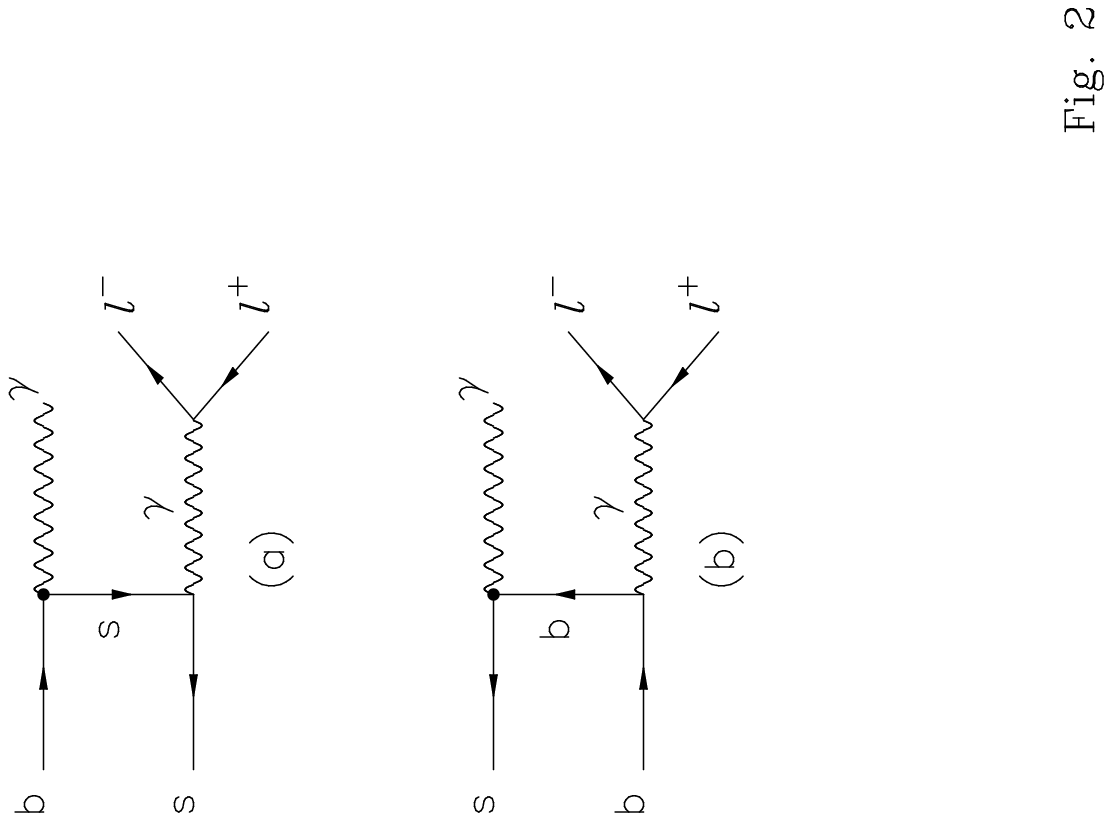}}
\end{figure}

\begin{figure}
\centerline{\epsffile{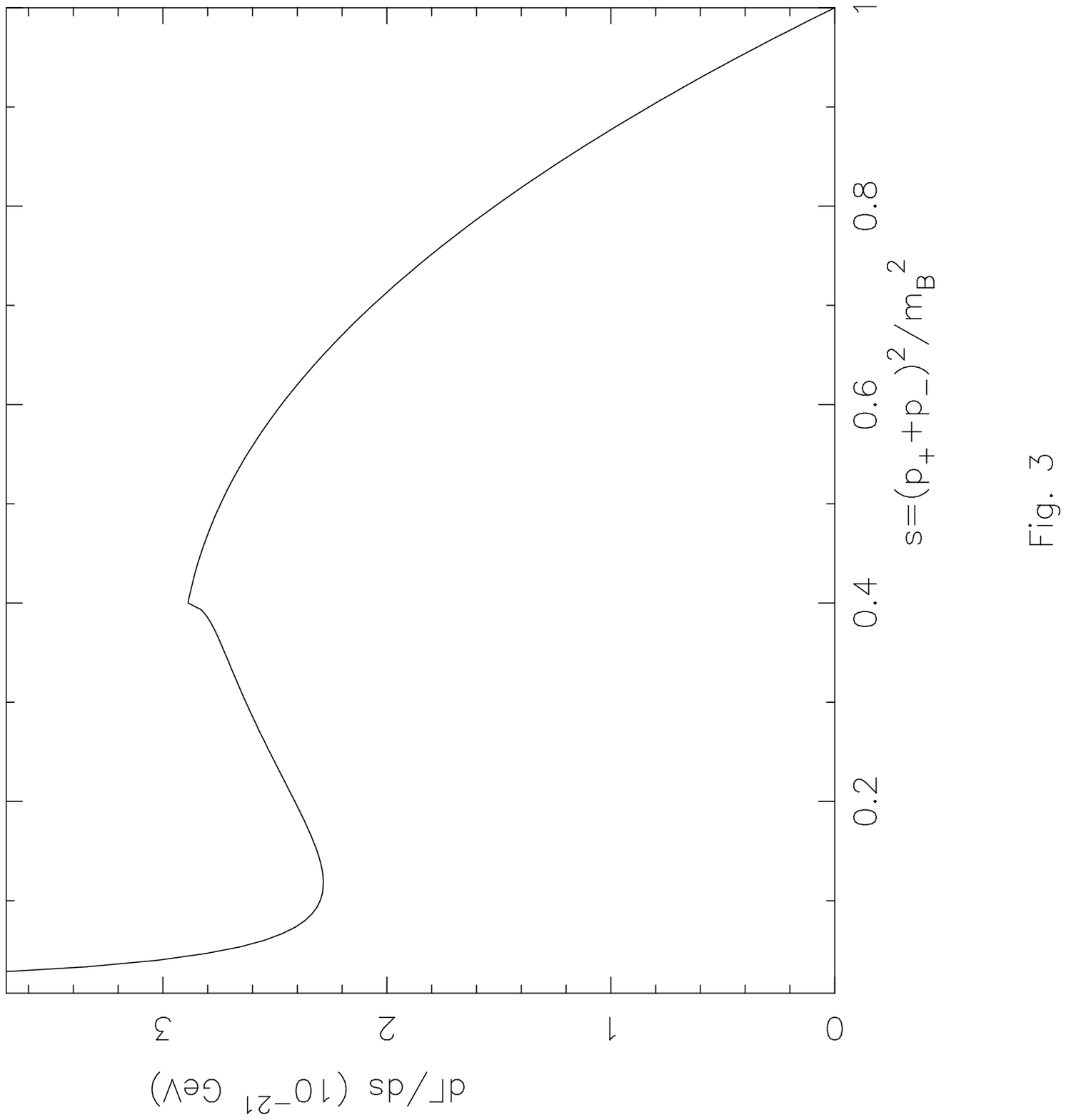}}
\end{figure}

\end{document}